\documentstyle[aps]{revtex} \draft \begin{document}
\title{FOKKER-PLANCK EQUATIONS FOR NUCLEATION PROCESSES 
REVISITED} \author{
David Reguera L\'opez,
J.M.  Rub\'{\i}, A.  P\'erez-Madrid} \address{Departament de F\'{\i}sica
Fonamental\\ Facultat de F\'{\i}sica\\ Universitat de Barcelona\\
Diagonal
647, 08028 Barcelona, Spain\\ }

\maketitle

\begin{abstract}

We present a new approach to analyze homogeneous nucleation based on 
non-equilibrium thermodynamics. The starting point is the formulation of a Gibbs 
equation for the variations of the entropy of the system, whose state is 
characterized by an internal coordinate or degree of freedom. By applying the 
method of non-equilibrium thermodynamics we then obtain the entropy production 
corresponding to a diffusion process in the internal space. The linear laws 
together with the continuity equation  lead to a kinetic 
equation of the Fokker-Planck type. By choosing properly the degree of freedom 
we are able to obtain a 
new kinetic equation for a global crystallization order parameter ( used in recent 
simulations ), and also we 
recover 
some of the existing equations. The consistency of the scheme we propose is 
proved in the quasi-stationary case. 
Finally, we also outline the way in which our 
formalism could be extended to more general situations.

\end{abstract}


\section{Introduction}
	
	The first step in many phase transformations of pure substances -for 
example condensation of gases or solidification of melts- is the formation of 
small embryos of the new phase within the bulk metastable substance. This is an 
activated process: a free energy barrier must be overcome in order to form 
embryos of a critical size, beyond which the new phase grows spontaneously. 
This 
fundamental mechanism of phase transformation is known as homogeneous 
nucleation, and the rate at which critical-sized embryos are formed is the 
nucleation rate ( for reviews on nucleation, see for example references  
\cite{kn:Kelton} - \cite{kn:debene} ).

	To analyze the process, three different types of 
schemes have basically been proposed in the literature.
The first one is referred to as classic nucleation theory.  In it, the process 
is usually described  by
means of a master equation that completely determines the evolution of the 
clusters 
size
distribution.  The fact that the size of the clusters vary in a discrete way 
(due to
the lost or gain of a single monomer) causes difficulties in the treatment
and resolution of this type of equation.  For this reason there have been
several trials to obtain a continuous diffusive equation as an approach to the
master equation.  The advantages that a continuous description may offer
are mainly the computational efficiency and the possibility of transforming the
Fokker-Planck equation into a Shr\"{o}dinger-type equation, soluble by means of
Quantum Mechanics' usual methods.
	The second type of description is centered on the construction of
dynamic models of the cluster interface.  By assuming that the molecules of the
interface behave as Brownian particles, it is possible to arrive at the 
characterization
of the clusters size distribution in terms of a Fokker-Planck equation.
	Finally, the last scheme is based on  the
theory of the density functional.  In this approach to the problem, the kinetic 
description  is carried out through the continuity equation for the
probability density.

	As we will see, the three points of view lead to express the nucleation
rate by means of a diffusive equation.  However, in spite of the great
variety of proposed equations, up to now none of them has turned out to be 
fully
satisfactory.

	In this paper we present an alternative approach that allows us to
obtain kinetic equations applicable to homogeneous nucleation. Our method 
is based on the formalism of internal degrees
of freedom \cite{kn:prigo}-\cite{kn:mazur}, formulated in the framework of 
non-equilibrium 
thermodynamics, and applied successfully before to 
other problems like the adsorption kinetics of particles on surfaces 
\cite{kn:absor}, the description of 
chemical 
reactions \cite{kn:pago} or the treatment of noise in interfaces of 
semiconductor devices \cite{kn:gabriel}. This formalism 
allows us to derive Fokker-Planck equations from a Gibbs 
equation
 and not as a  continuous approach to a master equation.  Moreover, within this 
framework we  recover directly the proper equilibrium distribution and we 
may 
derive explicit  expressions for drift and diffusion 
coefficients. 

	The paper is structured as follows.  In section II we will discuss the 
different procedures proposed (inside the scope of
homogeneous nucleation) to obtain Fokker-Planck  equations.  Section III is 
devoted to present our method  to derive a general Fokker-Planck equation 
describing the kinetics of homogeneous nucleation.  In section IV, we will 
particularize for several different
interpretations of the internal coordinate. We will derive a new kinetic 
equation in terms of an order parameter associated to the
degree of crystallization of the system and we will reproduce some of the
kinetic equations proposed in the literature.  In section V, we will discuss 
 the quasi-stationary case and show how in this case our approach overcomes 
some of the inherent 
difficulties of previous treatments.  Finally, in the last section we will 
stress our main results.

\section{ Brief review of previous treatments }

	In this section we will present a brief summary of the different 
Fokker-Planck
 equations that have been proposed to treat the problem of homogeneous 
nucleation.
  These equations have been obtained mainly from three different ways:  as
continuous approaches to a master equation; starting from microscopic
kinetic models by considering the separation of a molecule from the cluster 
surface like a problem of Brownian diffusion through a barrier; or as the
consequence of imposing a continuity equation for the probability density 
in the framework of the density functional theory. We will discuss all of
these
    in different subsections.

\subsection{ Continuous approaches to a master equation }

	In the classical model of nucleation \cite{kn:nuclea} one assumes 
that the size
of a cluster can vary due to the lost or absorption of a single monomer. Hence 
the evolution of the clusters distribution is described by the set of discrete
 master  equations

\begin{equation}\label{eq:i1} 
\frac{d \rho(n,t)}{dt}\,=\,\beta(n\!-\!1)\rho(n\!-\!1,t)\,+\, 
\alpha(n\!+\!1)\rho(n\!+\!1,t) 
\,-\,
\rho(n,t)\left ( \alpha(n)\,+\,\beta(n) \right )
\end{equation}  

\noindent where $\rho(n,t)$ represents the  density of clusters containing 
$n$ molecules and 
$\alpha(n)$ and $\beta(n)$ are  the rates at which a cluster of 
size $n$ loses or gains a 
single molecule, respectively.

	 To transform
this master equation into a continuous equation,
several different methods have been proposed in the literature.
	The most straightforward one consists of replacing the discrete 
variable $n$ 
by a continuous variable $x$.
 After expanding in Taylor series we arrive at the well-known Kramers-Moyal 
expansion,
 which upon truncation in second order yields the diffusion-like equation

\begin{equation}\label{eq:i2} \frac{\partial\rho}{\partial
t}\,=\,\frac{\partial}{\partial x}\left( D(x)\frac{\partial\rho}{\partial
x}\,+\,A(x)\rho \right)
\end{equation}

\noindent where $A(x)$ and $D(x)$ are  the drift and diffusion 
coefficients, respectively.

 This equation, however, has the  serious inconvenience that it does not 
reproduce
the equilibrium distribution of the original master equation. 
 Additionally, it  has also been
proven that it overestimates the nucleation rate exponentially, with an error
that grows unboundedly for large $n$ \cite{kn:hanggi84}.

	The Fokker-Planck equations of Frenkel-Zeldovich \cite{kn:frenkelj}, 
Goodrich 
\cite{kn:goodrich} and Shizgal-Barret \cite{kn:shizgal}  try to solve these 
difficulties by
imposing a relationship between
drift and diffusion coefficients in order to ensure the correct equilibrium
distribution. These three proposals differ in the expressions for those 
coefficients, 
chosen
to improve the approximation. However, as pointed out by Gitterman and Weiss 
\cite{kn:gitterman}, none of them seem to be able to reproduce the
correct deterministic behavior .

	A variant of these equations is the one formulated by Rabin and
Gitterman \cite{kn:rabin}, where the innovation  consist in considering
that the coefficients of the Fokker-Planck equation   cannot be fixed 
arbitrarily, but
rather - in the vicinities of the critical point - they are determined by
the critical dynamics of the system.

	Van Kampen \cite{kn:vankampen} proposed an alternative method, 
consisting 
in
rescaling the rate constants $\alpha$ and $\beta$ in terms of a parameter 
$\Omega$
(usually  the system size ) and to expand the result in powers of 
$\Omega^{-1}$.  In this way one obtains an expansion similar to Kramers-Moyal  
that
 shares the same problems described previously.

	The contribution of Grabert, H\"{a}nggi and Oppenheim\cite{kn:grabert}
 and later
H\"{a}nggi  et al. \cite{kn:shneidman} is also based on a rescaling of the
rate constants, but they obtain the coefficients (expressed in terms of the 
rates $\alpha$ and $\beta$) from 
 non-linear transport theory. 
Their equation reproduces the equilibrium distribution appropriately, but it
does not describe the short time behavior correctly  and it
gives an erroneous value of the variance of the distribution 
\cite{kn:gitterman}.

	Finally, inside the scope of  Fokker-Planck equation obtained starting
from the master equation, we can quote the  essentially mathematical method of 
Wu
\cite{kn:wu1},\cite{kn:wu2} that
proposes an index change and a mathematical optimization to minimize 
the error
associated to approach the  master equation - discrete - by means of a
 - continuous - diffusive type expression.

\subsection{ Kinetic Models }

	A different approach to obtain Fokker-Planck type equations 
for homogenous  nucleation, is based on the peculiar dynamics
 of a monomer near the surface of a cluster.  Inside this
scope we will consider the works of Lovett \cite{kn:lovett} and Ruckenstein and 
Narsimhan and 
Nowakowski \cite{kn:narshi} - \cite{kn:nowa}.

	Lovett uses Thermodynamics to estimate the radius of the critical
cluster  and the height of energy barrier that a monomer should surmount
 to leave the cluster surface.  Considering that the
monomer undergoes a Brownian movement, he uses the corresponding Fokker-Planck 
equation 
to study the probability of leaving the cluster.  His model 
 is only applicable to nucleation in gases, and he uses the macroscopic concept 
of interfacial tension 
 which can doubtfully be  applied to the
description of small clusters.

	Ruckenstein and coworkers \cite{kn:narshi} - \cite{kn:nowa} 
propose an alternative approach  based on 
the calculation of the rate $\alpha$. Their starting point is the estimation of 
the
potential barrier that is generated at the surface of a spherical cluster of 
radius $R$ from
the interaction between monomers and the recount of the  number of monomers
that interact with a given one.  They also consider that the movement of a
molecule is Brownian.  In the case of nucleation in liquids, they write 
a Smoluchowski equation  in spherical coordinates  for the distribution 
of probability $\rho(\vec{r},t)$

\begin{equation}\label{eq:i3} \frac{\partial\rho(\vec{r},t)}{\partial
t}\,=\,\vec{\nabla_{\vec{r}}}\cdot\left( D\vec{\nabla_{\vec{r}}}\,
\rho(\vec{r},t)
\,+\,\frac{D}{kT}\rho(\vec{r},t) \vec{\nabla_{\vec{r}}}\,\phi(\vec{r}) \right)
\end{equation}

\noindent Here $D$ is the diffusion coefficient and $\phi(\vec{r})$ the 
interaction 
potential of a surface monomer. For gases, they use a diffusion
 equation in the energy space. These equations are just employed to obtain the 
rates $\alpha$ and $\beta$. The rates are introduced in the common 
Kramers-Moyal truncated expansion of the master equation to evaluate the 
nucleation rate.

	The main inconveniences of this formalism are that it considers (in
the case of liquids) that the diffusion coefficient is constant, when in
fact there are reasons to think that it is not totally true; moreover the 
contribution of the bulk medium to the effective potential is not incorporated 
rigorously \cite{kn:debene}.  On the other hand, an important
difficulty linked to this type of descriptions arises from the fact that the 
nucleation rate is very
sensitive to uncertainties in the interatomic potential \cite{kn:oxtoby}.

\subsection { Density functional theory }

	One of the main drawbacks of the classic theory is the
assumption that the interface of the cluster is sharp,
hypothesis that - at least at the level of simulations -  is not
 completely satisfactory.

	Field models allow a treatment keeping in
mind the possibility that the interface between the cluster and the original
phase may be diffuse.
	The theory of the density functional \cite{kn:talan},\cite{kn:oxtoby} 
considers 
the
free energy as a functional of the density and uses it to built  a 
coarse-grained
free energy. The free
energy barrier can be taken out from variational principles. However,
 to be able to obtain kinetic information, hydrodynamic theory should 
be introduced.

	The starting point of Cahn and Hilliard kinetic treatment 
\cite{kn:cahn},\cite{kn:cahnH} 
is the formulation of an equation of
continuity for the current of probability density of a certain
configuration of the system. Following the scheme of non-equilibrium 
thermodynamics,
they postulate the following linear relationship 

\begin{equation}\label{eq:cahn} 
J\,=\,-M\nabla\frac{\partial F}{\partial \rho}
\end{equation}

\noindent where $J$ is the current, $M$ is the atomic mobility (taken as 
constant), $\rho$ is the probability density and $F$ is the coarse-grained
free energy. 
With these 
ingredients a  Fokker-Planck equation is obtained.

	The work of Langer \cite{kn:langer1} - \cite{kn:langer3}
constitutes an extension of the previous treatments in the sense that
it provides Statistical Mechanical basis where fluctuations
are taken into account, in the sense that he considers diffusion in the phase 
space. He proposes the following equation

\begin{equation}\label{eq:i4} \frac{\partial\rho(\{\eta\},t)}{\partial
t}\,=\,\sum_{i,j} \frac{\partial}{\partial \eta_{i}}\left( M_{ij} \left (
\frac{\partial\rho}{\partial \eta_{j}}\,+\,\frac{\partial G(\{\eta\})}{\partial 
\eta_{j}}
\frac{\rho}{kT} \right) \right)
\end{equation}	

\noindent where $\{\eta\}$ is the set of all degrees of freedom that 
characterize a system configuration, $G(\{\eta\})$ is the coarse - grained
free energy  functional, and $M_{ij}$ is a generalized mobility matrix.

	The main inconveniences of these theories based on the functional
density are that their correctness depend on the appropriate election of the
free energy functional and that its range of application  is
restricted; concretely they are not easily applicable to time dependent
situations.

\section{Non-equilibrium thermodynamics approach. Internal degrees of 
freedom}

In our approach to the problem of homogeneous nucleation, we consider 
that the state of a  system may be characterized by an 
internal coordinate or degree of freedom $\gamma$ , which may for example 
represent 
the number of monomers of a cluster (conveniently normalized to become a
continuous variable),
 its radius, or even an order parameter.
The nucleation process can be viewed as a diffusion process through a barrier, 
in the space spanned by the values of the internal coordinate, from an initial 
value $\gamma_{1}$ , corresponding to the metastable state, to $\gamma_{2}$ ,
characterizing the new phase.

Our starting point will be to assume that the variations of the entropy, 
$\delta S$ , due to the diffusion process are given by the Gibbs equation

\begin{equation}\label{eq:a1}
\delta S\,=\,-\,\frac{1}{T}\int_{\gamma_{1}}^{\gamma_{2}}\mu(\gamma,t)
\delta\rho(\gamma,t)d\gamma
\end{equation}

\noindent Here $\rho(\gamma,t)$ is the density in the internal space, 
$\mu(\gamma,t)$ its 
conjugated 
chemical potential and $T$ the temperature of the system which is assumed
 constant.
 
 In the absence of interactions among constituents of the system with different 
$\gamma$'s, the chemical 
potential is the 
 one for an ideal system \cite{kn:landau}

\begin{equation}\label{eq:a2}
\mu(\gamma,t)\,=\,k_{B}T\ln\rho(\gamma,t)\,+\,\Phi(\gamma)
\end{equation}

\noindent where $\Phi(\gamma)$ is the potential and $k_{B}$ the Boltzmann 
constant.

Quite generally, the evolution of the density in the internal space is 
governed by the continuity equation

\begin{equation}\label{eq:a3}
\frac{\partial\rho}{\partial t}\,=\,-\frac{\partial J}{\partial \gamma}
\end{equation}

\noindent which introduces the diffusion current in the internal space
$J(\gamma,t)$. The expression for the entropy production related to the
 diffusion process can be obtained from (\ref{eq:a1}) by using (\ref{eq:a3}).
 One then arrives at
 
\begin{equation}\label{eq:a4}
\sigma\,=\,-\frac{1}{T}\int J(\gamma)\frac{\partial\mu(\gamma)}{\partial 
\gamma}
d\gamma
\end{equation}

\noindent when a partial integration has been performed, with the assumption 
that the diffusion current vanishes at the initial and final states of the 
process.

Assuming locality in the internal space, for which only fluxes and forces with 
the same value of the internal coordinate are coupled, one then derives from 
expression 
(\ref{eq:a4}) the linear law

\begin{equation}\label{eq:a5}
J(\gamma)\,=\,-\frac{L(\gamma)}{T}\frac{\partial\mu(\gamma)}{\partial \gamma}
\end{equation}

\noindent when $L(\gamma)$ is a phenomenological coefficient, which may in 
general depend on the internal coordinate. This expression
can be used in the continuity equation (\ref{eq:a3}) thus arriving at the 
Fokker-Planck equation

\begin{equation}\label{eq:a6} \frac{\partial\rho}{\partial
t}\,=\,\frac{\partial}{\partial \gamma}\left( 
D(\gamma,t)\frac{\partial\rho}{\partial
\gamma}\,+\,b(\gamma,t)\frac{\partial\Phi}{\partial \gamma}\rho \right)
\end{equation}

\noindent where 

\begin{equation}\label{eq:b}
b(\gamma,t)\,=\,\frac{L(\gamma)}{\rho(\gamma,t) T }
\end{equation}

\noindent is a mobility in the internal space and

\begin{equation}\label{eq:a7} 
 D(\gamma,t)\,=\,k_{B}Tb(\gamma,t)
\end{equation}

\noindent the corresponding diffusion coefficient.

In order to fully determine our Fokker-Planck equation we need to specify the 
form of the diffusion coefficient in the internal space and the potential.
To obtain a microscopic explicit expression for 
the phenomenological 
coefficient $L(\gamma)$, we will apply fluctuating hydrodynamics in the 
internal space, in a similar way as in ref.\cite{kn:pago}. 
Adopting this framework, the presence of fluctuations in the diffusion process 
along the internal coordinate can be taken into account by adding a random 
contribution $J^{r}(\gamma,t)$ to the systematic contribution $J^{s}(\gamma,t)$ 
in the diffusion current

\begin{equation}\label{eq:a7b} 
 J(\gamma,t)\,=\,J^{s}(\gamma,t)\,+\,J^{r}(\gamma,t)
\end{equation}

\noindent Considering that $J^{r}(\gamma,t)$ has  zero mean and  satisfies the 
fluctuation-dissipation theorem

\begin{equation}\label{eq:fluct} 
 \langle J^{r}(\gamma,t)J^{r}(\gamma ',t ') 
\rangle\,=\,2k_{B}L(\gamma)\delta(\gamma  - \gamma ' )\delta(t  - t ')
\end{equation}

\noindent one can obtain the following  Green-Kubo 
expression for $L(\gamma)$

\begin{equation}\label{eq:gk} 
 L(\gamma)\,=\,\frac{1}{k_{B}} \int_{\gamma_{1}}^{\gamma_{2}}d \gamma ' 
\int_{0}^{\infty}dt \langle J^{r}(\gamma,0)J^{r}(\gamma ',t) \rangle
\end{equation}

\noindent Hence, in view of eqs.(\ref{eq:b}) and (\ref{eq:a7}) the diffusion
coefficient $D(\gamma,t)$ is given by

\begin{equation}\label{eq:D} 
 D(\gamma,t)\,=\,\frac{1}{\rho(\gamma,t)} \int_{\gamma_{1}}^{\gamma_{2}}d 
\gamma 
' 
\int_{0}^{\infty}dt \langle J^{r}(\gamma,0)J^{r}(\gamma ',t) \rangle
\end{equation}

\noindent This equation offers an explicit expression to determine 
the diffusion  coefficient in the internal space. For each interpretation
of the degree of freedom, the later equation could be transformed in order
to be evaluated by means of simulations. For instance, the way to calculate
the diffusion coefficient when the internal coordinate represents a global
order parameter will be extendedly explained in a future paper.

 Additionally, the nucleation barrier $\Phi(\gamma)$ can be 
estimated
by means of simulations in the way proposed by van Duijneveldt and Frenkel in 
ref. 
\cite{kn:barrera}. 
Consequently, our scheme becomes in this way completely soluble at least at the
simulation level.

Finally, it is worth pointing out that our formalism recover in a 
natural
and direct way ( without additional impositions ) the correct equilibrium
distribution. 
Since at equilibrium, the current in the internal space $J_{eq}$ vanishes, 
from equation 
(\ref{eq:a5}), this condition is accomplished
 when the chemical potential is constant. Using eq. (\ref{eq:a2}) we then
 obtain the 
proper equilibrium distribution

\begin{equation}\label{eq:a8} 
 \rho^{eq}(\gamma)\,\,\propto e^{ -\Phi(\gamma)/k_{B}T}
 \end{equation}

\section{Particular expressions of the Fokker-Planck equations.}

	The formalism we have proposed enables us to formulate a general 
Fokker-Planck equation in terms 
of  an arbitrary internal coordinate $\gamma$ specifying the state of the 
system. Thus, one of its inherent 
advantages  comes from the fact that 
it is not necessarily
restricted to give a vision of the nucleation only focused on the  
kinetics of clusters. Contrarily, a suitable choice of the internal coordinate
 will lead to the Fokker-Planck equation corresponding to a different 
description of the nucleation process.

 Our purpose in this section will 
be to particularize this equation for  different interpretations of the 
internal degree of freedom. In this way, we will see how we can propose a new 
Fokker-Planck equation for a global crystallization order parameter $Q$ and also 
how to reproduce the already 
existing ones.

\subsection { $\gamma$ represents an order parameter.}

 A possible way of characterizing the nucleation process, which has recently
been used in computer simulations (see ref. 
\cite{kn:frenkel},\cite{kn:frenkel2}),
  is by means of an 
order parameter 
 $Q$ that describes the global degree of 
 crystallization of the system.  In 
these simulations, the nucleation rate is calculated from a phenomenological 
rate equation using 
linear response theory.
According to our theory, the underlying Fokker-Planck equation describing 
the kinetics of this process in terms of the variable $Q$, considered as the
 internal degree of freedom, is given by

\begin{equation}\label{eq:b6} \frac{\partial\rho}{\partial
t}\,=\,\frac{\partial}{\partial Q}\left( D(Q,t)\frac{\partial\rho}{\partial
Q}\,+\,A(Q,t)\rho \right)
\end{equation}

\noindent  where, from eqs. (\ref{eq:a6})-(\ref{eq:a7}) the 
expressions of the drift and diffusion 
coefficients are

\begin{equation}\label{eq:Q1} 
 A(Q,t)\,=\,\frac{L(Q)}{T\rho(Q,t)}\frac{\partial\Phi(Q)}{\partial Q}
\end{equation}

\begin{equation}\label{eq:Q2} 
 D(Q,t)\,=\,\frac{k_{B}L(Q)}{\rho(Q,t)}
\end{equation}

	If the internal coordinate represents an order parameter 
associated to the whole 
crystalinity, the interpretation of $\rho$ is  slightly different.
 Now  $Q$  is a 
variable that characterizes the global state of the system. Therefore, the 
internal space associated to this variable would be the set of all the 
feasible replicas of the system in a given " macrostate ". From this point of 
view, $\rho$ 
would represent the non-equilibrium distribution of this ensemble whose 
meaning would be
the fraction of systems with value of the internal coordinate equal to $Q$ 
at time $t$. The quantity $\rho dQ$ would then be 
interpreted as the probability that the order parameter of our system (a single
realization of this 
non-equilibrium ensemble) 
has a value in the interval $(Q,Q+dQ)$ at time $t$.
Since in this case, we work with system replicas that obviously do not 
interact with each other, the system is ideal and eq. (\ref{eq:a2})
 for 
the chemical 
potential holds.

\subsection{ $\gamma$ related with the number of molecules in a cluster. }
 
	Let us assume that we choose as internal coordinate the number of 
monomers that constitute a cluster, $n$, or in a more convenient way, 
the dimensionless quantity $x\,=\,n/N$, with $N$ being the total number of
 monomers. In this case $\rho$ would represent the ( time-dependent ) 
 clusters size distribution, and $\Phi(x)$ would be the 
free energy needed to form a cluster of size $x$. Assuming that $x$ varies in 
a 
continuous way, we would recover the Fokker-Planck equation (\ref{eq:i2}) 
obtained as continuous extrapolation to a master equation,
with drift and diffusion coefficients formally analogous to 
expressions (\ref{eq:Q1}) and  (\ref{eq:Q2}).
	In this case, equation (\ref{eq:a2}) for the chemical potential
 will be
valid only if clusters do not interact. This lack of interaction is also 
implicitly
 assumed in previous treatments.

\subsection{ $\gamma$  is the  cluster radius. }

	If we consider the coordinate $\vec{r}$, corresponding to the position 
of a 
monomer in a cluster of radius $R$, as the internal degree of freedom and we 
work 
in spherical coordinates, we immediately recover the Smoluchowski equation 
proposed by 
Ruckenstein and co-workers \cite{kn:narshi}-\cite{kn:nowa}

\begin{equation}\label{eq:b4} \frac{\partial\rho(\vec{r},t)}{\partial
t}\,=\,\vec{\nabla_{\vec{r}}}\cdot\left( D(\vec{r},t)\vec{\nabla_{\vec{r}}}\,
\rho(\vec{r},t)
\,+\,\frac{D(\vec{r},t)}{kT}\rho(\vec{r},t) \vec{\nabla_{\vec{r}}}\,G(\vec{r}) 
\right)
\end{equation}

 \noindent Notice that now, since the phenomenological coefficient defined in 
eq. (\ref{eq:a5}) is in general a function of the internal coordinate, the 
diffusion coefficient will depend on the radial distance $\vec{r}$. In this case 
$\rho$ would represent the number density of monomers at $\vec{r}$, and $\phi$ 
the energy barrier to overcome in order to leave the cluster.

\subsection { $\gamma$ is the set of all system's degrees of freedom. }

	If instead of using a single coordinate  we carry out a complete 
description 
in terms of all the system's degrees of freedom $\{\eta\}$, characterizing a given 
configuration, we recover the form of the 
Fokker-Planck 
equation proposed by Langer \cite{kn:langer1} - \cite{kn:langer3}

\begin{equation}\label{eq:b5} \frac{\partial\rho(\{\eta\},t)}{\partial
t}\,=\,\sum_{i,j} \frac{\partial}{\partial \eta_{i}}\left( D_{ij} \left (
\frac{\partial\rho}{\partial \eta_{j}}\,+\,\frac{\partial\Phi}{\partial 
\eta_{j}}
\frac{\rho}{kT} \right) \right)
\end{equation}	

\noindent where now $D_{ij}\,=\,\frac{k_{B}L_{ij}(\{\eta\})}{\rho(\{\eta\},t)}$.
In this case, $\rho (\{\eta\}, t)$ corresponds to the probability 
density 
associated to the $\{\eta\}$ configuration
and $\Phi(\{\eta\})$ is the functional  free energy.

\section { The quasi-stationary case }

In the previous section we have seen how  to reproduce some
 different forms of the Fokker-Planck 
equations proposed in the literature. Our task now will be to ascertain if the
equations  we have obtained, not only have the appropriate form, 
but are also physically consistent.  To this purpose,
we will focus our attention on the clusters size distribution equation.
	
Most of the proposed Fokker-Planck equations are just different continuous 
diffusive approximations to a discrete master equation. Thus, the rightness of 
this kind 
of equations is evaluated in terms of the accuracy of 
the approximation. In particular, the two main requirements 
that a Fokker-Planck equation must satisfy is that 
it reproduces the equilibrium distribution and the deterministic growth
rate $\dot{n}$ \cite{kn:shneidman}; in other words, it must describe nucleation 
and growth 
appropriately.

The Fokker-Planck equation we present has been derived directly and not as 
a continuous approach to a master equation. Therefore, in general we do not 
have an underlying master equation model to which our expression should be 
adjusted and that serve us as a criterion to judge its accuracy.

However, we have shown that by construction our equation always guarantees the correct 
equilibrium 
distribution. Moreover, a case exists in which we can build a master equation  
to evaluate the correctness of our results. 
It is the quasi-stationary case, corresponding to high nucleation 
barriers.

 When the height of the nucleation barrier is large enough as compared with 
thermal 
 energy, the system achieves a quasi-stationary state characterized by the 
current
	
\begin{equation}\label{eq:d1} 
J(n,t)\,=\,J(t) \left\{ \theta(n-n_{1})\,-\,
\theta(n-n_{2}) \right\}
\end{equation}

\noindent This form implies that equilibrium is reached independently at each 
barrier side, consequently the chemical potential will be uniform  
 
\begin{equation}\label{eq:d2} 
\mu(n,t)\,=\,\mu(n_{1},t)\theta(n_{0}-n)\,+\,
 \mu(n_{2},t)\theta(n-n_{0})
\end{equation}
  
\noindent Here $n_{0}$ specifies the size of the critical cluster.  
  
	Substituting these two equations in (\ref{eq:a5}) and integrating 
with respect to the coordinate $n$, we  obtain 
  
 \begin{equation}\label{eq:d3} 
J(t)\,=\,\frac{D(n_{0})}{n_{2}\,-\,n_{1}} \left ( \rho(n_{1}
,t) \exp{ \left( -\frac{ \Phi(n_{0})-\Phi(n_{1}) }{k_{B}T} \right )}\,-
\,\rho(n_{2}
,t) \exp{  \left (-\frac{\Phi(n_{0})-\Phi(n_{2}) }{k_{B}T} \right ) }
\right )
\end{equation}

	If the nucleation barrier is high enough in such a way that a 
quasi-stationary state is 
reached, 
the clusters distribution of intermediate sizes does not vary in time, 
so that only the states corresponding to $\rho(n_{1},t)$ and 
$\rho(n_{2},t)$ matter. The value $n_{1}$ represents the smaller cluster 
distinguishable from equilibrium fluctuations in the metastable phase and 
$n_{2}$ 
corresponds to a postcritical stable cluster. Clusters sizes $n_{1}$  
and $n_{2}$
( where $n_{1}<n_{0}<n_{2}$ ) are chosen such that for $n<n_{1}$, 
$\rho(n)=\rho_{eq}(n)$ and for 
$n>n_{2}$, 
$\rho(n)=0$. These boundary conditions are habitual in classical treatments 
\cite{kn:Kelton}
 and final results do not strongly depend on the values of 
 $n_{1}$
and $n_{2}$ \cite{kn:greer}. If the intermediate sizes 
 distribution  is constant, it implies that the lost of one $n_{1}$ cluster
  supposes,
  after jumping the energy barrier, the formation of one postcritical nucleus. 
  Conversely, the gain of a $n_{1}$ cluster is due to the 
disappearance of a postcritical cluster.  
  
	Accordingly, the master equation  for this quasi-stationary
 situation has the following form:
 
 \begin{equation}\label{eq:me1} 
\frac{d \rho(n_{1},t)}{dt}\,=\,k^{-}\rho(n_{2},t)\,-\,k^{+}\rho(n_{1},t)
 \,=\,- J(t)
\end{equation}   
 
\begin{equation}\label{eq:me2} 
\frac{d \rho(n,t)}{dt}\,=\,0 ,\ \ \ \ \  n_{1}<n<n_{2}
\end{equation}   
 
 \begin{equation}\label{eq:me3} 
\frac{d \rho(n_{2},t)}{dt}\,=\,k^{+}\rho(n_{1},t)\,-\,k^{-}\rho(n_{2},t)
 \,=\, J(t)
\end{equation}   
 
\noindent where $k^{+}$ and $k^{-}$ are the forward and backward rates. 
The forward rate $k^{+}$ is simply the probability that a cluster $n_{1}$ 
disappears, or equivalently, 
that it surpasses the nucleation barrier. Therefore, $k^{+}$ will be proportional 
to the Boltzmann factor 
associated to the height of the barrier

\begin{equation}\label{eq:d7} 
k^{+}\,=\,\lambda \exp{\left ( -\frac{ \Phi(n_{0})-\Phi(n_{1})}{k_{B}T} \right ) 
}
\end{equation}   

\noindent Similarly, for the rate $k^{-}$ one has  
 
 \begin{equation}\label{eq:d8} 
 k^{-}\,=\,\lambda \exp{ \left ( -\frac{\Phi(n_{0})-\Phi(n_{2}) }{k_{B}T} \right 
) }
\end{equation}   

\noindent Notice that the constant $\lambda$ enters both expressions of $k^{+}$ 
and $k^{-}$  
in order to
 guarantee that in equilibrium the flow $J_{eq}$ vanishes. This is the requirement
 to fulfill detailed balance. 

Therefore, the nucleation rate $J(t)$ obtained from the master equation  
is given by
 
  \begin{equation}\label{eq:d9} 
J(t)\,=\,\lambda \left ( \rho(n_{1}
,t) \exp{ \left( -\frac{ \Phi(n_{0})-\Phi(n_{1}) }{k_{B}T} \right ) }\,-
\,\rho(n_{2}
,t) \exp{ \left (-\frac{\Phi(n_{0})-\Phi(n_{2}) }{k_{B}T} \right ) }
\right )
\end{equation}
  
	\noindent Identifying $\lambda$ with  the factor 
$\frac{D(n_{0})}{n_{2}\,-
\,n_{1}}$ in eq. (\ref{eq:d3}), we see that at least in the case of high
 nucleation barriers, our diffusive equation reproduces the steady-state 
 nucleation rate.    	
Moreover, we are going to prove that our formalism also recovers all the
 distribution moments of the 
master equation model.

By introducing (\ref{eq:d1}) into the continuity equation (\ref{eq:a3}) 
 we obtain the following expression

\begin{equation}\label{eq:continu}
\frac{\partial\rho(n,t)}{\partial t}\,=\,- J(t) \left\{ \delta(n-n_{1})\,-\,
\delta(n-n_{2}) \right\}
\end{equation}

\noindent which yields the evolution equation for the $r$-moment 
($r=1,2,3\ldots$)

\begin{equation}\label{eq:fpmoment} 
\frac{d\!<\!n^{r}\!>}{dt}\,=\,(n_{2}^{r} - n_{1}^{r})J(t) 
\end{equation}   

\noindent It is immediate to realize that this expression agrees
 with the corresponding one evaluated from the set 
of master equations (\ref{eq:me1})-(\ref{eq:me3}).

  Therefore, we have proved the validity of our equation in the 
quasi-stationary
   case, in the sense that it 
 satisfies the two rightness criteria required to make our approach consistent,
 and moreover reproduces the same results obtained from a 
   master equation. 
	If the height of the barrier is low, the 
quasi-stationary hypothesis is not longer valid. This feature invalidate the 
considerations we have made in this section.

\section {Conclusions}

In this paper we have reviewed the different Fokker-Planck equations
 proposed in the literature to treat the problem of  homogeneous
nucleation.  We have seen that the equations obtained as continuous
 approximation to a master equation
  have the main drawback of not to be able to
reproduce some of the master equation characteristics (as the equilibrium
distribution, the deterministic growth, the variance of the distribution or the 
short
time behavior). Moreover, in all of them (except  in the vicinities of the 
critical
 point )
the coefficients are not fully determined, but rather 
their expression are postulated, or they are left in terms of the unspecified 
rates $\alpha$
 and $\beta$. To estimate these rates an appropriate microscopic kinetic 
model 
is required.

	On the other hand,  equations obtained from kinetic models
base their validity on the correction of the microscopic model proposed
for the dynamics of the cluster interface. They do not give a
complete and correct explicit expression for drift  and diffusion coefficients.
	Finally, the field-theory approach   is restricted to the
vicinity of the critical point  and to situations where temporal
dependencies do not exist.

	 We have proposed a new  method based upon non-equilibrium 
thermodynamics in the space of an internal coordinate characterizing the state 
of the system. This method allows us to easily obtain a set of general kinetic
equations of the Fokker-Planck type that not only
  reproduce the ones which have been proposed, centered on the evolution of
clusters size  distribution, but  they allow alternative
descriptions of the nucleation process. In this context, we have derived a new 
Fokker-Planck equation in terms of the crystallinity parameter $Q$ used by 
Frenkel et al. \cite{kn:frenkel},\cite{kn:frenkel2} in recent simulations.

	We have verified that the Fokker-Planck equation we have obtained 
apart 
from having
the appropriate form, it fulfills the two validity 
criteria required for this type of equations,
at least in the quasi-stationary case. In fact, the quasi-stationary case 
corresponds to low supersaturations ( and slow rates ), that are the typical 
experimentally affordable 
conditions.

	Our expression also keeps in mind all the possible dependencies (as 
much 
in
$n$ as in $t$), and it is able to propose a  microscopic expression to 
calculate ( by means  simulations )
the drift and diffusion coefficients.  

	Our model allows to treat time-dependent nucleation and also
effects of a preexistent clusters distribution. 
	Moreover, this new proposed formalism, not only reproduces and 
overcomes some of the
 main drawbacks of previous homogeneous nucleation treatments, but rather
  it constitutes an 
 appropriate framework to treat nucleation processes from a wider perspective.
   In this context, we
could take into account the possible interactions among clusters just by
introducing an
activity factor in the expression of the chemical potential
(\ref{eq:a2}).
	Moreover, the formalism could easily be 
extended to the case of heterogeneous nucleation and
even to analyze hydrodynamic effects in the nucleation
process by simply introducing  additional variables in the Gibbs equation 
characterizing the liquid 
phase  \cite{kn:agustin},\cite{kn:mazur}.

\acknowledgments

We would like to thank Prof. H. Reiss for fruitful discussions. 
This work has been supported by DGICYT of the Spanish Government under
grant PB95-0881.  D.  Reguera wishes to thank Generalitat de
Catalunya for financial support.

\end{document}